\documentclass[prl,twocolumn,letterpaper,showpacs,groupedaddress,superscriptaddress,nofootinbib,floatfix,preprintnumbers,tightenlines]{revtex4-1}

\usepackage[colorlinks,urlcolor=blue]{hyperref}
\usepackage{amsmath,amssymb,amsfonts}
\usepackage{graphicx}
\usepackage{bm}
\usepackage{comment}
\usepackage{color}
\usepackage{csquotes}
\usepackage{longtable}
\usepackage[section]{placeins}

\allowdisplaybreaks
\let\Oldsection\section
\renewcommand{\section}{\FloatBarrier\Oldsection}
\let\Oldsubsection\subsection
\renewcommand{\subsection}{\FloatBarrier\Oldsubsection}

\newcount\hour \newcount\hourminute \newcount\minute
\hour=\time \divide \hour by 60
\hourminute=\hour \multiply \hourminute by 60
\minute=\time \advance \minute by -\hourminute
\newcommand{\mydate}{\ \today \ - \number\hour :\number\minute}

\def\OMIT#1{{}}

\def\si{^1 \hskip -0.03in S _0}
\def\siii{^3 \hskip -0.025in S _1}

\def\diii{^3 \hskip -0.03in D _1}

\newcommand{\overbar}[1]{\mkern 1.9mu\overline{\mkern-1.9mu#1\mkern-1.9mu}\mkern 1.9mu}

\begin{document}

\title{
Entanglement Suppression and  Emergent Symmetries of Strong Interactions }

\preprint{INT-PUB-18-056}
\preprint{NT@UW-18-19}

\author{Silas R.~Beane}
\affiliation{Department of Physics, University of Washington, Seattle, WA 98195-1560, USA}

\author{David B.~Kaplan}
\affiliation{Institute for Nuclear Theory, University of Washington, Seattle, WA 98195-1550, USA}

\author{Natalie Klco}
\affiliation{Department of Physics, University of Washington, Seattle, WA 98195-1560, USA}
\affiliation{Institute for Nuclear Theory, University of Washington, Seattle, WA 98195-1550, USA}

\author{Martin J.~Savage}
\affiliation{Institute for Nuclear Theory, University of Washington, Seattle, WA 98195-1550, USA}

\date{\mydate}

\begin{abstract}
  \noindent
  Entanglement suppression in the strong interaction $S$-matrix is
  shown to be correlated with approximate spin-flavor symmetries that
  are observed in low-energy baryon interactions, the Wigner $SU(4)$
  symmetry for two flavors and an $SU(16)$ symmetry for three flavors.
  We conjecture that dynamical entanglement suppression is a property
  of the strong interactions in the infrared, giving rise to these
  emergent symmetries and providing powerful constraints on the nature
  of nuclear and hypernuclear forces in dense matter.
\end{abstract}
\maketitle

Understanding approximate global symmetries in the strong interactions
has played an important historical role in the development of the
theory of Quantum Chromodynamics (QCD).  Baryon number symmetry arises
 in QCD because it is impossible to include a marginal or relevant
interaction consistent with Lorentz and gauge symmetry which violates
baryon number, while the axial and vector flavor symmetries are
understood to be due to the small ratio of quark masses (and their
differences) to the QCD scale.  The approximate low-energy $SU(2n_f)$
spin-flavor symmetry for $n_f=2,3$ flavors which relates spin-1/2 and
spin-3/2 baryons can be understood as arising at leading order (LO) in the
large-$N_c$ expansion, where $N_c$ is the number of
colors~\cite{tHooft:1973alw,Witten:1980sp}.  In low-energy nuclear
physics, a different spin-flavor symmetry is observed in the structure
of light-nuclei and their $\beta$-decay rates, namely Wigner's $SU(4)$
symmetry, where the two spin states of the two nucleons transform as
the 4-dimensional fundamental
representation~\cite{Wigner:1936dx,Wigner:1937zz,Wigner:1939zz}.  It
has been shown that this symmetry also arises from the large-$N_c$
expansion at energies below the $\Delta$ mass
\cite{Kaplan:1995yg,Kaplan:1996rk,CalleCordon:2008cz}.  The agreement
of large-$N_c$ predictions with nuclear phenomenology has been
extended to higher-order
interactions~\cite{Liu:2017otd,Schindler:2018irz,Savage:1995kv,Polinder:2006zh},
three-nucleon
systems~\cite{Phillips:2013rsa,Epelbaum:2014sea,Vanasse:2016umz}, and
to studies of hadronic parity
violation~\cite{Phillips:2014kna,Schindler:2015nga,Savage:2000iv}. Recently,
however, lattice QCD computations for $n_f=3$ have revealed an
emergent $SU(16)$ symmetry in low-energy interactions of the baryon
octet---analogous to Wigner's $SU(4)$, but with the two spin states of the eight baryons transforming as the 16-dimensional representation of
$SU(16)$~\cite{Wagman:2017tmp}.  This low energy symmetry has been
lacking an explanation from QCD. In this Letter, we show that both
Wigner's $SU(4)$ symmetry for $n_f=2$ and $SU(16)$ for $n_f=3$
correspond to fixed lines of minimal quantum entanglement in the
$S$-matrix for baryon-baryon scattering, and we propose entanglement
suppression to be a dynamical property of QCD and the origin of these
emergent symmetries~\footnote{
A principle of {\it maximum} entanglement
  has been previously proposed to constrain quantum electrodynamics in
  Ref.~\cite{Cervera-Lierta:2017tdt}.
  }.

Of the many features of quantum mechanics and quantum field theory
(QFT) that dictate the behavior of subatomic particles, entanglement and
its associated non-locality are perhaps the most striking in their
contrast to everyday experience.  The degree to which a system is
entangled, or its deviation from tensor-product structure, provides a
measure of how ``non-classical'' it is.  The importance of
entanglement as a feature of quantum theory has been known since the
work of Einstein, Podolsky and Rosen~\cite{Einstein:1935rr} and later
pioneering papers \cite{Bell:1964kc,Aspect:1981zz,Freedman:1972zza},
and has become a core ingredient in quantum information science,
communication and perhaps in understanding the very fabric of
spacetime~\cite{Ryu:2006bv,Ryu:2006ef,Maldacena:2013xja}.  Despite this
long history, the implications of entanglement in QFTs, e.g.,
Refs.~\cite{Buividovich:2008gq,Buividovich:2008yv,Nakagawa:2009jk,Donnelly:2011hn,Casini:2013rba,Radicevic:2014kqa,Ghosh2015,Itou:2015cyu,Aoki2015,Soni:2015yga,PhysRevLett.117.131602,Witten:2018lha},
and in particular for experimental observables in high-energy and
heavy-ion collisions are only now starting to be
explored~\cite{Rogers:2010dm,Ho:2015rga,Kharzeev:2017qzs,Berges:2017zws,Shuryak:2017phz,Baker:2017wtt,Berges:2017hne,Hagiwara:2017uaz,Liu:2018gae,Kovner:2018rbf,Cervera-Lierta:2017tdt}. Here
we study the role of entanglement in low-energy nuclear interactions.

In general, a low-energy scattering event can
entangle position, spin, and flavor quantum numbers, and it is
therefore natural to assign an entanglement power to the $S$-matrix
for nucleon-nucleon scattering.  We choose to define the entanglement
power of the $S$-matrix in a two-particle spin
space~\cite{PhysRevA.63.040304,mahdavi2011cross}, noting that this
choice is not unique and that others will be explored elsewhere
\cite{TBA}.  This is determined by the action of the $S$-matrix on an
incoming two-particle tensor product state with randomly-oriented
spins, $|\psi_\text{in}\rangle = \hat R(\Omega_1)
|\hspace{-0.3em}\uparrow \rangle_1 \otimes \hat R(\Omega_2)
|\hspace{-0.3em}\uparrow \rangle_2 $, where $ \hat R(\Omega_j)$ is the
rotation operator acting in the $j^{\rm th}$ spin-${1\over 2}$ space,
and all other quantum numbers associated with the states have been
suppressed.
For low-energy processes, this random spin pair projects
onto the two states with total spin $S=0,1$,
and associated phase shifts $\delta_{0,1}$, in the
$\si$ and $\siii$ channels, respectively, with projections onto higher
angular momentum states suppressed by powers of the nucleon momenta.
The entanglement power, ${\mathcal E}$, of the $S$-matrix, $\hat {\bf
  S}$, is defined as
\begin{equation}
{\mathcal E}({\hat {\bf S}})  = 1- \int {d\Omega_1\over 4\pi} \ {d\Omega_2\over 4\pi}\  {\rm Tr}_1\left[ \ \hat\rho_1^2 \  \right]
\ \ \ ,
\label{eq:epgen}
\end{equation}
where $\hat\rho_1 = {\rm Tr}_2\left[\ \hat\rho_{12}\ \right]$ is the
reduced density matrix for particle 1 of the two-particle density matrix
$\hat\rho_{12} = |\psi_\text{out}\rangle\langle \psi_\text{out}| $ with $|\psi_\text{out} \rangle =
\hat {\bf S}|\psi_\text{in}\rangle$.
By describing the average action of $\hat {\bf S}$ to transition a tensor-product state to an entangled state,
the entanglement power expresses a state-independent entanglement measure that vanishes when   $|\psi_\text{out}\rangle$ remains a tensor product state for any   $|\psi_\text{in}\rangle$.

Following the analysis of
Ref.~\cite{Cervera-Lierta:2017tdt}, we consider
the spin-space entanglement of two distinguishable particles, the
proton and neutron for $n_f=2$ QCD.
Neglecting the small tensor-force-induced mixing of the $\siii$ channel with the $\diii$ channel,
the $S$-matrix for low-energy scattering below inelastic threshold in these sectors can be decomposed as
\begin{eqnarray}
  \hat {\bf S}
  & = &
{1\over 4}\left( 3 e^{i 2 \delta_1} + e^{i 2 \delta_0} \right)
 \hat   {\bf 1}
\ +\
{1\over 4}\left( e^{i 2 \delta_1} - e^{i 2 \delta_0} \right)
  \hat  {\bm \sigma} \cdot   \hat  {\bm \sigma}
 ,
  \label{eq:Sdef}
\end{eqnarray}
where
$ \hat {\bf 1} = \hat {\cal I}_2\otimes  \hat {\cal I}_2$
and
$  \hat {\bm \sigma} \cdot \hat {\bm \sigma} = \sum\limits_{\alpha=1}^3 \
\hat{\bm \sigma}^\alpha \otimes \hat{\bm \sigma}^\alpha$.
It follows that the entanglement power of $ \hat {\bf S}$ is
\begin{eqnarray}
{\mathcal E}({\hat {\bf S}})
& = &
{1\over 6}\ \sin^2\left(2(\delta_1-\delta_0)\right)
\ \ \   ,
  \label{eq:epSnf2}
\end{eqnarray}
which vanishes when $\delta_1-\delta_0= m {\pi\over 2}$ for any
integer $m$.  This includes the $SU(4)$ symmetric case
$\delta_1=\delta_0$ where the coefficient of $\hat {\bm \sigma} \cdot
\hat {\bm \sigma}$ vanishes.  Special fixed
points where the entanglement power vanishes occur when the phase
shifts both vanish, $\delta_1=\delta_0=0$, or are both at unitarity,
$\delta_1=\delta_0=\frac{\pi}{2}$, or when
$\delta_1=0$, $\delta_0=\frac{\pi}{2}$ or $\delta_1=\frac{\pi}{2}$,
$\delta_0=0$. The S-matrices at these fixed points with vanishing
entanglement power are $\hat {\bf S}=\pm \hat {\bf 1}$ and
$\pm ( \hat {\bf 1}+ \hat {\bm \sigma} \cdot \hat
{\bm \sigma})/2$~\footnote{The S-matrices at the four fixed points realize
a representation of the Klein four-group, $\mathbb{Z}_2\otimes \mathbb{Z}_2$.}.

\begin{figure}
  \includegraphics[width=0.95\columnwidth]{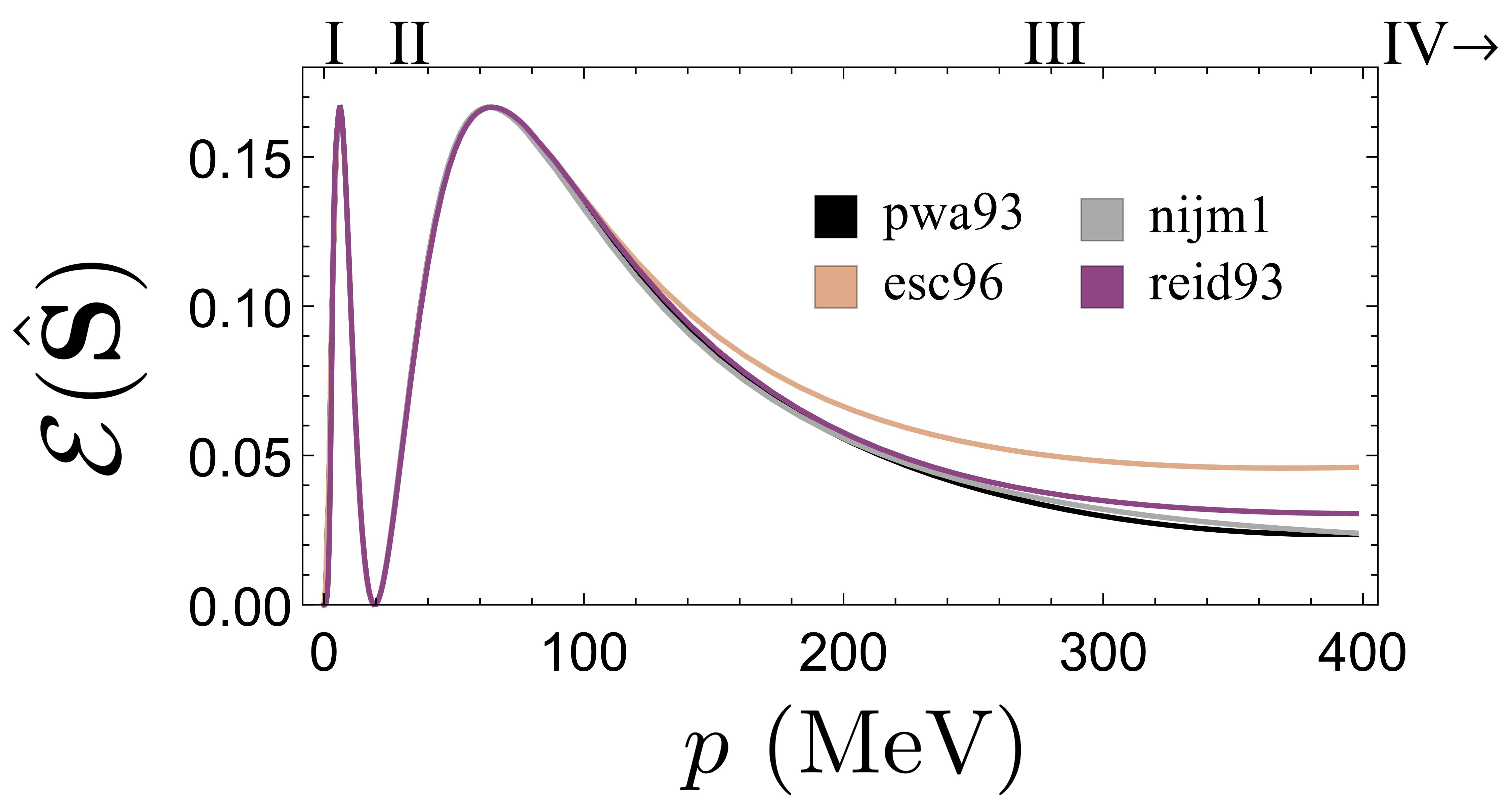}
  \caption{The entanglement power,
    $\mathcal{E}(\hat{\mathbf{S}})$, of the $S$-matrix as a
    function of $p$, the center-of-mass nucleon momentum.  The $\si$
    and $\siii$ phase shifts used to calculate
    $\mathcal{E}(\hat{\mathbf{S}})$ were taken from four
    different
    models~\cite{PhysRevC.48.792,PhysRevC.49.2950,PhysRevC.54.2851,PhysRevC.54.2869,NNOnline}
    to provide a na\"ive estimate of systematic uncertainties. Data for this figure may be found in Table~\ref{tab:fig1mt} in the supplemental material. }
  \label{fig:epPW}
\end{figure}

The entanglement power in nature is plotted in Fig.~\ref{fig:epPW} as
a function of the center-of-mass nucleon momentum, $p$, up to pion
production threshold, making use of the $\si$ and $\siii$ phase shifts
derived from the analyses of
Refs.~\cite{PhysRevC.48.792,PhysRevC.49.2950,PhysRevC.54.2851,PhysRevC.54.2869}.
The four regions indicated are distinguished by the role of
non-perturbative physics.  Region I shows that entanglement
power approaches zero in the limit $p\rightarrow 0$, as will be the
case for any finite range interaction not at unitarity.  At momenta
around the scale of the inverse scattering lengths, region II, poles
and resonances of $\hat{\mathbf{S}}$ produce highly-entangling
interactions.
This non-perturbative
structure could be considered a source of ultra-low-momentum
entanglement power; experimental evidence for this is expected to be
found in the vanishing modification of $np$-scattering quantum correlations at
19.465(42)~MeV where the phase shifts differ by $\pi/2$ and
 $|p \hspace{-0.3em} \uparrow , n \hspace{-0.3em}\downarrow\rangle$
 scatters into
$|p \hspace{-0.3em}\downarrow  , n\hspace{-0.3em} \uparrow\rangle$.
In region IV, where energies are of order the chiral symmetry breaking scale,
the entangling interactions of
quark and gluon degrees of freedom become prominent.
It is region III
that is the main focus of this paper---away from the far-infrared
structure but with nucleons as fundamental degrees of freedom, the
entanglement power is suppressed.
Once relativistic corrections and
$\siii$-$\diii$ mixing---parametrically suppressed at low-energy---are
included in Eq.~\eqref{eq:Sdef},
$\mathcal{E}(\hat{\mathbf{S}})$ is expected to remain
suppressed but non-zero, indicating that the entanglement suppression
in nature is only partial.

Much progress has been made in nuclear physics in recent years by
considering  low-energy effective field theories (EFTs), constrained by data from
nucleon scattering.
The $\delta_{0,1}$ phase shifts can be computed for energies below the
pion mass, from the pionless EFT for
nucleon-nucleon interactions. The leading interaction in the effective
Lagrangian is
\begin{equation}
{\cal L}^{n_f=2}_{\rm LO}
=
-\frac{1}{2} C_S (N^\dagger N)^2
-\frac{1}{2} C_T \left(N^\dagger{\bm \sigma}N\right)\cdot \left(N^\dagger{\bm \sigma}N\right) \ ,
\label{eq:interaction}
\end{equation}
where $N$ represents both spin states of the proton and neutron
fields. These interactions can be re-expressed as contact interactions
in the $\si$ and $\siii$ channels with couplings $\overbar{C}_0 = ( C_S-3 C_T) $
and $\overbar{C}_1 = (C_S+C_T)$ respectively, where the two couplings are fit to
reproduce the $\si$ and $\siii$ scattering lengths.
The $\overbar{C}$
coefficients both run with the renormalization group as described in Ref.~\cite{Kaplan:1998tg,Kaplan:1998we}
with a stable IR fixed point at
$\overbar{C}=0$, corresponding to free particles, and a nontrivial, unstable IR
fixed point at $\overbar{C}=C_\star$ corresponding to a divergent scattering
length and constant phase shift of $\delta=\pi/2$ (the ``unitary" fixed
point).
At the four fixed points (described above), where $\{\overbar{C}_0, \overbar{C}_1\}$ take the values
$0$ or $C_\star$, the theory has a conformal
(``Schr\"odinger")
symmetry; there is also a fixed line of enhanced symmetry at $C_T=0$,
or equivalently $\overbar{C}_0=\overbar{C}_1$, where the theory possesses the Wigner
$SU(4)$ symmetry, as apparent from the form of
Eq.~(\ref{eq:interaction}) with $C_T=0$.  When fitting to the
scattering lengths one finds $C_T\ll C_S\simeq C_\star$, since
scattering lengths are unnaturally large in both channels.  Therefore,
low-energy QCD has approximate $SU(4)$ symmetry and sits close to the
$\{C_\star,C_\star\}$ conformal fixed point~\cite{Mehen:1999qs}. The
emergence of $SU(4)$ symmetry (but not necessarily conformal symmetry)
follows from the large-$N_c$ expansion where  $C_T/C_S =
O(1/N_c^2)$~\cite{Kaplan:1995yg}.
\begin{figure}[t]
  \includegraphics[width=0.8\columnwidth]{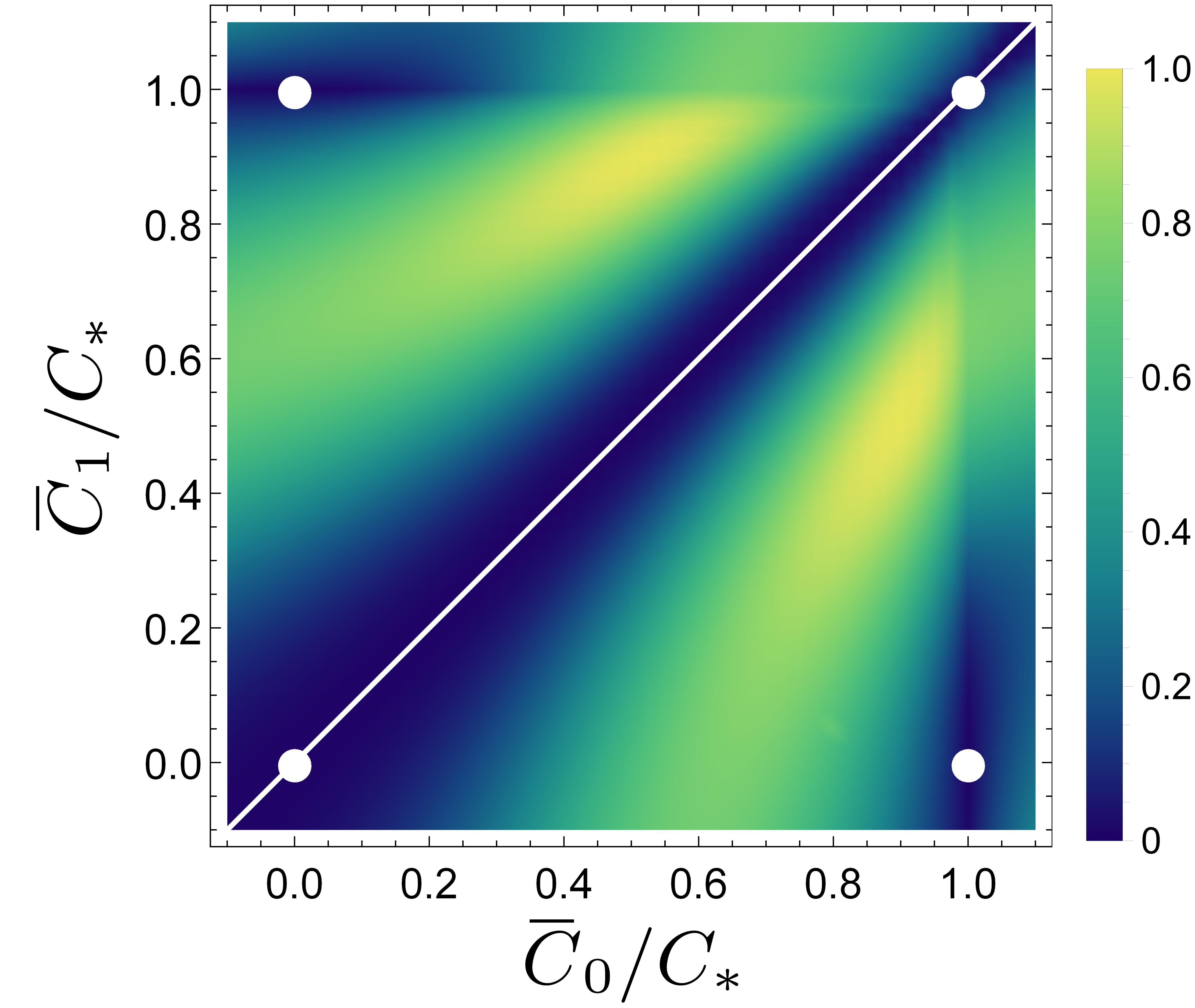}
  \caption{
  Density plot of the entanglement power
    $\mathcal{E}(\hat{\mathbf{S}})$ of the $S$-matrix (see Eq.~\eqref{eq:SmatEP} of the supplemental material) integrated
    over center of mass momenta $0\le p\le m_\pi/2$, versus the
    Lagrangian couplings $\overbar{C}_0/C_\star$ and $\overbar{C}_1/C_\star$ where
    $C_\star$ is the critical coupling for unitary scattering.
    The entanglement power vanishes
    at the four conformal fixed-points (white points), as
    well as the fixed line corresponding to Wigner $SU(4)$ symmetry
    (white diagonal).
    }
  \label{fig:ES}
\end{figure}

The symmetry points of the EFT can be related to
minimization of the entanglement power of the $S$-matrix.
Fig.~\ref{fig:ES} shows a density plot of ${\mathcal E}({\hat {\bf
    S}})$ as computed from Eq.~(\ref{eq:interaction}) averaged
over momenta $0\le p \le m_\pi/2$, as  a function of the couplings
$\overbar{C}_{0,1}$ renormalized at $\mu = m_\pi/2$ and rescaled by $C_\star$.
Superimposed in white are the four conformal fixed points, as well as
as the Wigner $SU(4)$ fixed line.
The minima of the
entanglement power of the $S$-matrix (${\cal E}(\hat {\bf S}) = 0$)  coincide with the points of enhanced
symmetry in the EFT; the $SU(4)$ line corresponds
to $\delta_0=\delta_1$ for all momenta, while the conformal points off
the $SU(4)$ line correspond to $|\delta_0-\delta_1|=\pi/2$.

In the $n_f=2$ case, the large-$N_c$ expansion gives a similar
expectation for $SU(4)$ symmetry as does a principle of entanglement
suppression. However, an analogous equivalence does not hold for $n_f=3$,
as the large-$N_c$ expansion predicts the conventional approximate
$SU(6)$ spin-flavor symmetry, while entanglement suppression predicts a much
larger $SU(16)$ symmetry under which the two spin states of the baryon
octet transform as a 16-dimensional representation.  To see this,
consider the EFT in the $SU(3)$ flavor symmetry limit of
QCD, where  six independent contact operators
contribute at LO~\cite{Savage:1995kv},
\begin{eqnarray}
\mathcal{L}_{\rm LO}^{n_f=3}
& = &
-c_1 \langle B_i^{\dagger}B_iB_j^{\dagger}B_j \rangle
-c_2  \langle B_i^{\dagger}B_jB_j^{\dagger}B_i\rangle
\nonumber\\
& &
  -c_3  \langle B_i^{\dagger}B_j^{\dagger}B_iB_j\rangle
-c_4  \langle B_i^{\dagger}B_j^{\dagger}B_jB_i\rangle
\nonumber\\
& &
 -c_5  \langle B_i^{\dagger}B_i\rangle   \langle B_j^{\dagger}B_j\rangle
-c_6  \langle B_i^{\dagger}B_j\rangle   \langle B_j^{\dagger}B_i\rangle
\ ,
\label{eq:BB-Lagrangian}
\end{eqnarray}
where $\langle ... \rangle$ denotes  a trace in flavor space,
and $B_i$ is the $3\times 3$ octet-baryon matrix
where the subscript $i=1,2$ denotes spin.
$\mathcal{L}_{\rm LO}^{n_f=3}$ is invariant  under rotations and the
transformation $B\rightarrow VBV^\dagger$ where $V$ is an $SU(3)$ matrix.
In the large-$N_c$ limit of QCD, an $SU(6)$ spin-flavor symmetry emerges relating the six
coefficients $c_i$ in Eq.~(\ref{eq:BB-Lagrangian}) to two independent
coefficients $a, b$~\cite{Kaplan:1995yg} in the $SU(6)$ invariant
Lagrange density,
\begin{eqnarray}
c_1  =  -{7\over 27} b
\ ,\
c_2\  &=& \ {1\over 9} b
\ ,\
c_3\ = \ {10\over 81} b
\ ,\
\nonumber \\
c_4\ = \ -{14\over 81} b
\ ,\
c_5\ &=& \ a\ +\ {2\over 9} b
\ ,\
c_6\ = \ -{1\over 9} b
    \  .
\label{eq:cisu6}
\end{eqnarray}
A comprehensive set of lattice QCD calculations of light nuclei,
hypernuclei and low-energy baryon-baryon scattering in the limit of
$SU(3)$ flavor symmetry by the NPLQCD
collaboration~\cite{Beane:2012vq,Beane:2013br,Wagman:2017tmp}
demonstrates that the $c_i$ are consistent with this predicted $SU(6)$
spin-flavor symmetry~\cite{Wagman:2017tmp}.  The two-baryon sector
calculated with $m_\pi\sim 800~{\rm MeV}$ is found to be
unnatural~\cite{Beane:2012vq,Beane:2013br,Wagman:2017tmp}, with a
scattering length that is larger than the range of the interaction,
and hence better described by the power-counting of van
Kolck~\cite{vanKolck:1998bw} and
KSW~\cite{Kaplan:1998tg,Kaplan:1998we,Chen:1999tn}.  Further, the
values of $c_1, c_2, c_3, c_4$ and $c_6$ are calculated to be much
smaller than $c_5$, indicating that $b \ll
a$~\cite{Beane:2012vq,Beane:2013br,Wagman:2017tmp}.  When $b=0$, the
$SU(6)$ is enlarged to an emergent $SU(16)$ spin-flavor
symmetry~\cite{Wagman:2017tmp}, where the baryon states populate the
fundamental of $SU(16)$,
{
\medmuskip=0mu
\thinmuskip=0mu
\thickmuskip=0mu
\begin{eqnarray}
\mathcal{L}_{\rm LO}^{n_f=3} & \rightarrow &  -{1\over 2} c_S \left({\cal B}^\dagger {\cal B}\right)^2
 \ \ \
{\cal B}\ =\  \left( p_\uparrow , p_\downarrow , n_\uparrow , n_\downarrow, \Lambda_\uparrow , ... \right)^T
 , \ \ \
\label{eq:su16LO}
\end{eqnarray}
}
with $c_S=2 c_5$.

The existence of $SU(16)$ symmetry and $b=0$ does not follow from the
large-$N_c$ expansion,
but does follow from entanglement suppression.
The  entanglement power of the $S$-matrix in spin-space from
the $n_f=3$ interactions in Eq.~(\ref{eq:BB-Lagrangian})
can be addressed
by
considering its action on states of distinguishable baryons.
Computing the entanglement power ${\mathcal E}({\hat {\bf S}})$ for
more than six distinct two-baryon channels with nonidentical
particles---e.g., $\Lambda N$, $\Xi^- p$---shows  that
zero entanglement power occurs at the $SU(16)$ point
where all the $c_n$ couplings vanish except for $c_5$, which is
unconstrained (and all LO scattering matrices in the $J=0$ and $J=1$
mixed-flavor sectors are
diagonal~\cite{Savage:1995kv,Wagman:2017tmp}).
Thus, the
principle of entanglement suppression gives rise to an approximate
symmetry, apparent in lattice QCD
calculations~\cite{Beane:2012vq,Beane:2013br,Wagman:2017tmp},  that
does not follow from the large-$N_c$ limit.
We conclude that  the
large-$N_c$ limit of QCD does not provide a sufficiently stringent
constraint to produce a low-energy EFT that does not entangle, which
could not be deduced from the $n_f=2$ sector alone~\cite{Kaplan:1995yg}.
Thus, the
entanglement power of the $S$-matrix
appears to be a dominant ingredient in dictating
the properties and relative size of interactions in low-energy nuclear
and hypernuclear systems.

While in nuclei and hypernuclei contributions to binding from
 three-body forces between nucleons
and hyperons are small compared with those from  two-baryon
forces, they cannot be neglected and become more important with
increasing density.
To understand whether entanglement suppression dictates approximate $SU(16)$ symmetry in these interactions as well,
we take a more general approach rather than computing the multi-baryon $S$-matrix in various channels to constrain couplings.
We begin by assuming exact
 $SU(2)_\text{spin} \times SU(3)_\text{flavor}$
symmetry, where corrections due to $SU(3)$ violation from quark mass differences can be incorporated in the usual way.
Even in the degenerate quark mass limit, this means restricting ourselves to considering only interactions that do not couple spin to orbital angular momentum.
While such spin-orbit and tensor interactions can be important in heavy nuclei, they are suppressed by powers of the baryon momenta and do not enter the IR limit of the effective theory.
It is then argued that entanglement suppression requires the interactions to respect a $U(1)^{16}$ symmetry,   conserving particle number individually  for each of the octet baryon spin states.
To see why this is a reasonable assumption,
consider a 1-body operator (which need not be local) that violates the $U(1)^{16}$ symmetry, e.g.,
\begin{equation}
\medmuskip=2.5mu
\thinmuskip=2.5mu
\thickmuskip=2.5mu
\hat\Theta\
=\
\int \text{d}^3 {\bf v}  \text{d}^3 {\bf u}\
\left[
f({\bf v}  - {\bf u}) \alpha_{{\bf v} }^\dagger \beta_{{\bf u}}   + {\rm h.c.}
\right]   ,
\end{equation}
where $\alpha, \beta$ are annihilation operators for components of $\mathcal{B}$ with
$\alpha \neq \beta$, ${\bf u}$ and ${\bf v}$ are spatial coordinates and $f$ is a form factor.
This operator implements the transformation, e.g.,
\begin{eqnarray}
 \hat\Theta |\alpha_{\bf x}, \beta_{\bf y}, \gamma_{\bf z}\rangle  =
    \int \text{d}^3{\bf w} \
    && \big[
    f({\bf w}-{\bf y}) |\alpha_{\bf x}, \alpha_{\bf w}, \gamma_{\bf z}\rangle
\nonumber\\
&&
    \ +\
     f^*({\bf x}-{\bf w}) |\beta_{\bf w}, \beta_{\bf y}, \gamma_{\bf z}\rangle
     \big]
   ,
\end{eqnarray}
producing an entangled state, even if $f({\bf x}-{\bf y}) = \delta^3({\bf x}-{\bf y})$,
from which it can be concluded
 that the $U(1)^{16}$ symmetry is a necessary condition to forbid entangling
 interactions~\footnote{
 The converse is not true: it is possible to show that there exist entangling interactions which preserve  $U(1)^{16}$ symmetry \cite{TBA}.
 }.
 It follows  from simultaneous exact $SU(2)\times SU(3)$ and $U(1)^{16}$ symmetries  that the
 LO EFT must respect the full $SU(16)$ symmetry by the following argument.    The charges $Q_\alpha = \mathcal{B}^\dagger\Gamma_\alpha \mathcal{B}$  that by assumption commute with the Hamiltonian $H$ consist of
\begin{equation}
  \Gamma_\alpha \in \left\{\mathcal{I}_{16},
  \ S_i \otimes \mathcal{I}_8,\ \mathcal{I}_2 \otimes t_a,\ M_i \right\}
  \ \ \ ,
  \label{eq:gensi}
\end{equation}
where $S_{1,2,3} \in \mathfrak{su}(2)$ are the fundamental
generators of $SU(2)$, $ t_a \in \mathfrak{su}(3)$ with $(t_a)_{bc} = -i f_{abc}$ for
$a, b, c = 1,..., 8$ are the generators of the $SU(3)$ adjoint
representation with structure constants $f_{abc}$, and the $M_i$ for $i =
1,..., 15$ are a set of independent diagonal traceless $16\times 16$
matrices generating $U(1)^{15}$, the ignored $U(1)$ symmetry being baryon number.
Since all of the above $Q^\alpha$ are assumed to commute with $H$, it follows that their commutators do as well.
The full symmetry of $H$ will be the symmetry group   generated by the closure of the $Q^\alpha$ under commutation.
By making use of the fact that the $t_a$ generate an irreducible representation of the $\mathfrak{su}(3)$ Lie algebra and invoking Schur's Lemma, it is possible to show that   this full symmetry algebra is $\mathfrak{su}(16)$ \cite{TBA}.

Conjecturing that the guiding principle for low-energy nuclear and
hypernuclear forces is the suppression
of entanglement fluctuations provides important theoretical constraints on dense matter systems.
The Lagrange density describing the $n_f=2$ sector
with vanishing entanglement power, and therefore $SU(4)$ spin-flavor
symmetry, is
\begin{eqnarray}
{\cal L}^{(n_f=2)}  & = &
-\sum_{n = 2}^{4}{1\over n!} C^{(n)}_S \left(N^\dagger N\right)^n
\ \ \ ,
\label{eq:su4zero}
\end{eqnarray}
while for $n_f=3$ with $SU(16)$ spin-flavor symmetry,
\begin{eqnarray}
{\cal L}^{(n_f=3)}  & = &
-\sum_{n = 2}^{16}{1\over n!} c^{(n)}_S \left({\cal B}^\dagger {\cal B}\right)^n
 .
\label{eq:su16zero}
\end{eqnarray}
Calculations of hypernuclei and hyperon-nucleon interactions imposing
$SU(16)$ spin-flavor symmetry on the low-energy forces are now in
progress~\cite{su16nuclei}.  Our work suggests that such calculations
could probe the nature of entanglement in strong interactions.

The Pauli exclusion principle's requirement of anti-symmetrization
produces a natural tendency for highly entangled states of identical
particles in the $s$-channels.  It is somewhat perplexing how to
understand the result that the $S$-matrix for baryon-baryon scattering exhibits screening of
entanglement power when the quarks and gluons that form the nucleon
are highly entangled.  It may be the case that the nonperturbative
mechanisms of confinement and chiral symmetry breaking together
strongly screen entanglement fluctuations in the low-energy sector of
QCD beyond what can be identified in the large-$N_c$ limit of QCD.

While our  work has focused  on low-energy interactions,
preliminary evidence for entanglement suppression at higher orders in a derivative expansion
is seen in the $n_f=2$ low-energy constants (LECs) for operators up to NNLO.
The contact terms of the two-nucleon potential in the center-of-mass frame are~\cite{Epelbaum:2004fk}
\begin{eqnarray}
  V_{\rm contact} &=& C_S  + C_T \, \vec{\sigma}_1 \cdot  \vec{\sigma}_2 + V^{(2)}_{\rm contact}~,\\
V^{(2)}_{\rm contact} &=& C_1 \, \vec{q}\,^2 +
 C_3 \, \vec{q}\,^2  ( \vec{\sigma}_1 \cdot \vec{\sigma}_2) + C_6 \, (\vec{q}\cdot \vec{\sigma}_1 )(\vec{q}\cdot \vec{\sigma}_2 )
~, \nonumber
\end{eqnarray}
with $\vec{q} = \vec{p}\hspace{0.2em}'-\vec{p}$ and $\vec{p}, \vec{p}\hspace{0.2em}'$ the
initial and final nucleon momenta.
Calculating their entanglement power, it is expected that $C_T, C_3,$ and
$C_6$ will be suppressed at low energies.
Numerical values of these potential coefficients are determined from the values of
the spectroscopic
LECs~\cite{Carlsson:2015vda,Epelbaum:2001fm,Epelbaum:2014efa}
(see Fig.~1 of the supplementary material).
At small
values of the maximum scattering energy, $T^\text{max}_\text{Lab}$, the coefficients of the non-entangling operators,
$C_S$ and
$C_1$, are found to be larger in magnitude than their entangling counterparts.
Furthermore, as $T^\text{max}_\text{Lab}$ is increased and
shorter distances scales are probed, the suppression lessens and $C_6$
grows.
While these observations are consistent with
entanglement-suppressed LECs, work
remains to be done in understanding
the mechanism that suppresses entanglement power
in the transition from QCD
to low-energy effective interactions, and the full consequences of this mechanism.

Nuclear physics, with its rich theoretical structure
and phenomenology emerging from QCD and QED in the infrared, provides a unique forum for the study of fundamental properties
of quantum entanglement.
We conjecture that the suppression of entanglement is an important element of strong-interaction physics
that is correlated with enhanced emergent  symmetries.

\vskip 0.2in
	We would like to thank A.  Cervera-Lierta, D. Lonardoni, A. Murran, and A. Roggero for inspiring discussions and A. Ekstr\"om for providing additional details for the LEC correlations in Ref.~\cite{Carlsson:2015vda}.
	SRB was supported in part by the U. S. Department of Energy grant DE-SC001347.
	DBK, NK and MJS  were supported  by DOE grant No.~DE-FG02-00ER41132,
	and NK was supported in part by a Microsoft Research PhD Fellowship.


\bibliography{bibelenp}

\clearpage
\onecolumngrid

\begin{center}
\large\textbf{Supplemental Material for\\ \enquote{Entanglement Suppression and Emergent Symmetries of Strong Interactions}}
\end{center}
\vspace{1cm}

\begin{figure}[h]
  \centering
  \includegraphics[width=0.48\textwidth]{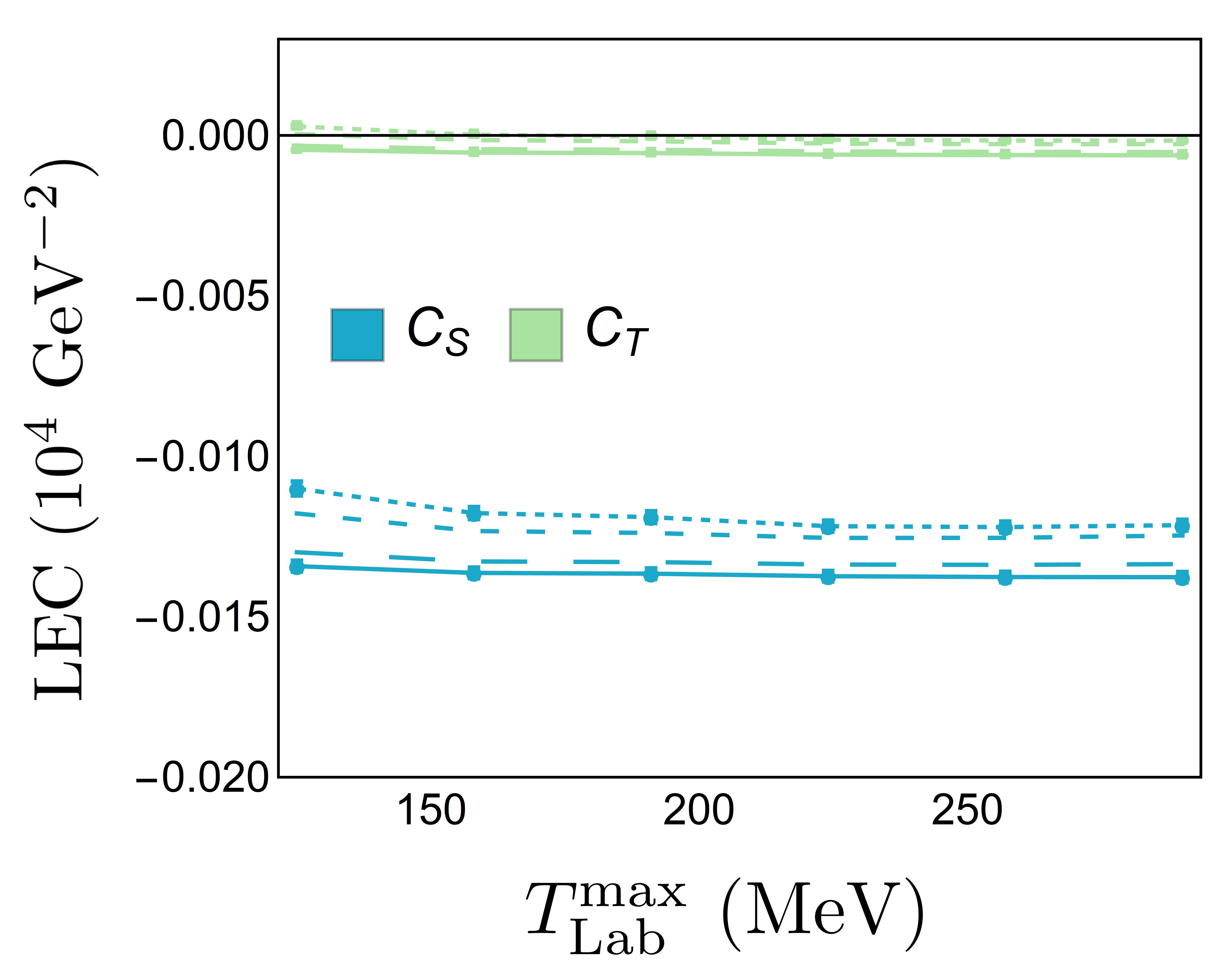}
  \hspace{0.3cm}
  \includegraphics[width=0.48\textwidth]{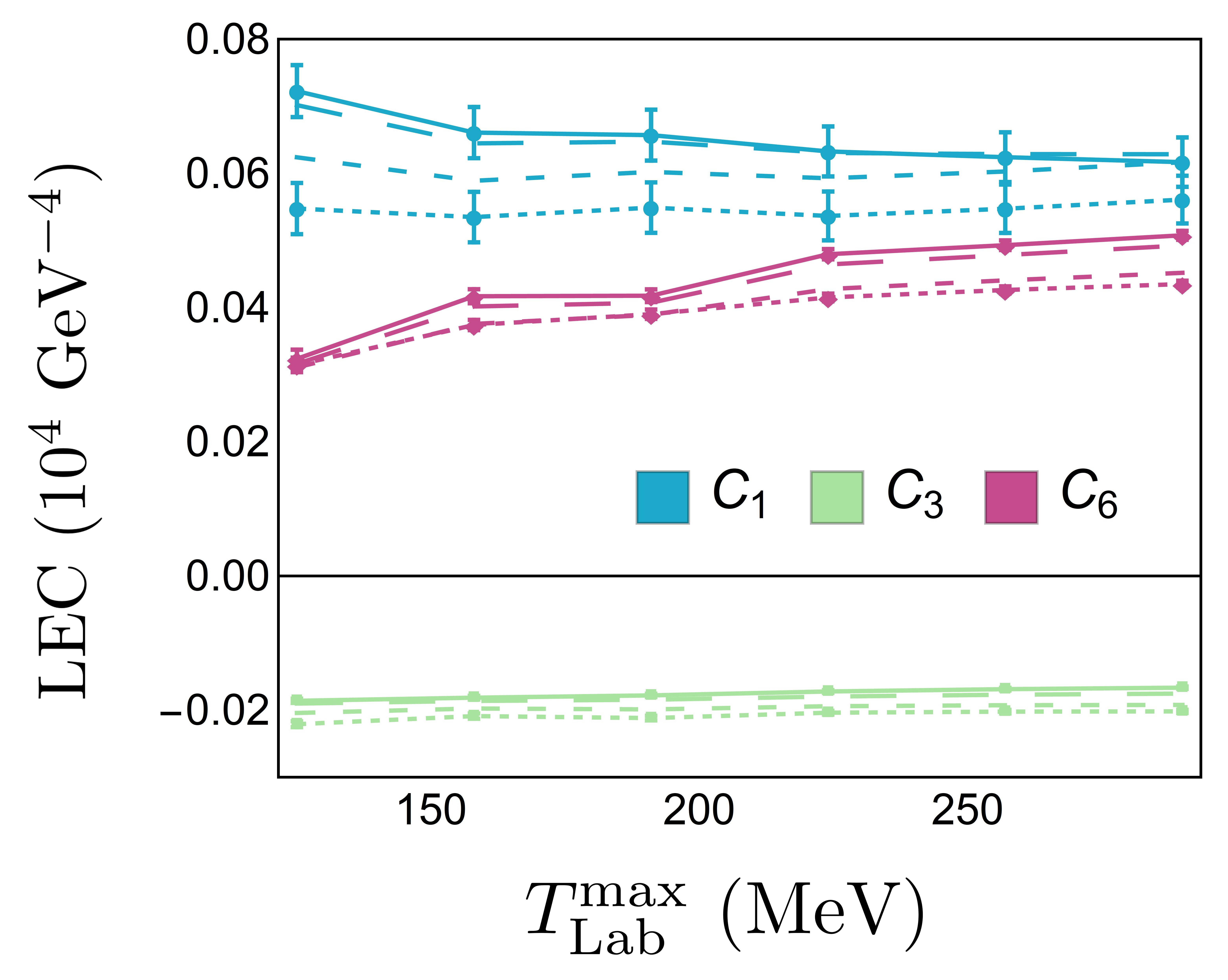}
  \caption{
  Low-energy constants of the NNLO two-nucleon interaction at $\mathcal{O}\left(p^0\right)$ (left panel)
  and $\mathcal{O}\left(p^2\right)$  (right panel)
  as a function of the maximum scattering energy retained in the fitted
  experimental data set, $T_{\text{Lab}}^\text{max}$,
  determined from NN and $\pi$N scattering~\cite{Carlsson:2015vda}.
  The value of the regulator cutoff corresponds to
  $\Lambda$ = 450~MeV for the solid line,
  $\Lambda$ = 475~MeV for the long-dashed line,
  $\Lambda$ = 550~MeV for the dashed line and
  $\Lambda$ = 600~MeV for the dotted line.
  The uncertainties on the solid and dotted lines are propagated from the correlated uncertainties of the spectroscopic LECs fit in Ref.~\cite{Carlsson:2015vda}.
  }
  \label{fig:LECs}
\end{figure}

In this supplemental material, Fig.~\ref{fig:LECs} shows the scaling of low energy constants (LECs) relevant for the two-nucleon potential in the center-of-mass frame.
The data in Fig.~\ref{fig:LECs} have been compiled from the spectroscopic LECs and associated correlated uncertainties
that were fit in Ref.~\cite{Carlsson:2015vda}.  With the inclusion of experimental data up to a maximum scattering energy $T_\text{Lab}^\text{max}$, the stability of the potential coefficients suggests that the LECs are well-constrained by low-energy data well-below pion threshold.
The progression from solid to dashed to dotted lines in  Fig.~\ref{fig:LECs} shows the shift in these coefficients with increasing regulator-energy cutoffs of
$\Lambda = 450, 475, 550, 600$ MeV.
Numerical values of the data shown in Fig.~\ref{fig:LECs} may be found in Table~\ref{tab:fig2data} while that of Fig.~1 of the main text may be found in Table~\ref{tab:fig1mt}.
\setlength\LTleft{-30pt}%
\setlength\LTright{-30pt}
\begin{center}
\begin{longtable}{c|cccccc}
\hline
\hline
$\Lambda = 450 $ & $T_{\text{Lab}}^{\text{max}} =  $125 & $T_{\text{Lab}}^{\text{max}} =  $158 & $T_{\text{Lab}}^{\text{max}} =  $191 & $T_{\text{Lab}}^{\text{max}} =  $224 & $T_{\text{Lab}}^{\text{max}} =  $257 & $T_{\text{Lab}}^{\text{max}} =  $290 \\
\hline
$C_S$ & -0.01342(17) & -0.01363(17) & -0.01366(17) & -0.01374(17) & -0.01376(17) & -0.01377(16) \\
$C_T$ & -0.00045(2) & -0.00053(2) & -0.00055(2) & -0.00060(2) & -0.00061(1) & -0.00062(1) \\
$C_1$ & 0.0723(39) & 0.0661(38) & 0.0657(38) & 0.0633(37) & 0.0624(37) & 0.0617(37) \\
$C_3$ & -0.0186(3) & -0.0181(3) & -0.0178(2) & -0.0172(2) & -0.0169(2) & -0.0167(2) \\
$C_6$ & 0.0324(14) & 0.0417(10) & 0.0418(9) & 0.0480(8) & 0.0493(7) & 0.0508(7) \\
\hline
$\Lambda = 475 $ & $T_{\text{Lab}}^{\text{max}} =  $125 & $T_{\text{Lab}}^{\text{max}} =  $158 & $T_{\text{Lab}}^{\text{max}} =  $191 & $T_{\text{Lab}}^{\text{max}} =  $224 & $T_{\text{Lab}}^{\text{max}} =  $257 & $T_{\text{Lab}}^{\text{max}} =  $290 \\
\hline
$C_S$ & -0.01299(18) & -0.01328(17) & -0.01330(17) & -0.01337(17) & -0.01338(17) & -0.01336(16) \\
$C_T$ & -0.00032(2) & -0.00043(2) & -0.00045(2) & -0.00050(2) & -0.00051(2) & -0.00052(2) \\
$C_1$ & 0.0702(39) & 0.0645(38) & 0.0647(38) & 0.0630(37) & 0.0629(37) & 0.0629(37) \\
$C_3$ & -0.0190(3) & -0.0186(3) & -0.0184(2) & -0.0180(2) & -0.0177(2) & -0.0175(2) \\
$C_6$ & 0.0316(13) & 0.0402(10) & 0.0407(9) & 0.0464(7) & 0.0478(7) & 0.0493(6) \\
\hline
$\Lambda = 550 $ & $T_{\text{Lab}}^{\text{max}} =  $125 & $T_{\text{Lab}}^{\text{max}} =  $158 & $T_{\text{Lab}}^{\text{max}} =  $191 & $T_{\text{Lab}}^{\text{max}} =  $224 & $T_{\text{Lab}}^{\text{max}} =  $257 & $T_{\text{Lab}}^{\text{max}} =  $290 \\
\hline
$C_S$ & -0.01178(19) & -0.01233(18) & -0.01239(18) & -0.01254(17) & -0.01254(17) & -0.01247(17) \\
$C_T$ & 0.00004(4) & -0.00014(3) & -0.00018(3) & -0.00025(2) & -0.00027(2) & -0.00027(2) \\
$C_1$ & 0.0624(38) & 0.0589(38) & 0.0602(37) & 0.0593(37) & 0.0603(36) & 0.0617(36) \\
$C_3$ & -0.0204(4) & -0.0198(3) & -0.0199(3) & -0.0194(2) & -0.0193(2) & -0.0192(2) \\
$C_6$ & 0.0311(11) & 0.0376(8) & 0.0389(7) & 0.0428(6) & 0.0441(6) & 0.0452(5) \\
\hline
$\Lambda = 600 $ & $T_{\text{Lab}}^{\text{max}} =  $125 & $T_{\text{Lab}}^{\text{max}} =  $158 & $T_{\text{Lab}}^{\text{max}} =  $191 & $T_{\text{Lab}}^{\text{max}} =  $224 & $T_{\text{Lab}}^{\text{max}} =  $257 & $T_{\text{Lab}}^{\text{max}} =  $290 \\
\hline
$C_S$ & -0.01100(22) & -0.01177(19) & -0.01189(19) & -0.01218(18) & -0.01220(17) & -0.01214(17) \\
$C_T$ & 0.00028(5) & 0.00002(3) & -0.00003(3) & -0.00013(3) & -0.00016(3) & -0.00016(2) \\
$C_1$ & 0.0548(38) & 0.0535(38) & 0.0549(38) & 0.0537(36) & 0.0547(36) & 0.0561(36) \\
$C_3$ & -0.0221(4) & -0.0209(4) & -0.0212(3) & -0.0204(3) & -0.0202(3) & -0.0202(2) \\
$C_6$ & 0.0315(11) & 0.0374(8) & 0.0390(7) & 0.0415(6) & 0.0426(5) & 0.0435(5) \\
\hline
\hline
    \multicolumn{7}{l}{} \\[-0pt]
    \caption{
    Numerical values of the data shown in Fig.~\ref{fig:LECs}. Both $\Lambda$ and $T_{\text{Lab}}^{\text{max}}$ are given in MeV.
    The values of $C_{S,T}$ and $C_i$ are given in units of $10^{4}$ GeV$^{-2}$ and $10^4$ GeV$^{-4}$, respectively.
    Central values and correlated uncertainties are calculated from the spectroscopic LECs fit  to NN and $\pi$N scattering data given in Ref.~\cite{Carlsson:2015vda}.
    }
    \label{tab:fig2data}
\end{longtable}
\end{center}

\begin{figure}[!ht]
\centering
\includegraphics[width = 0.8\textwidth]{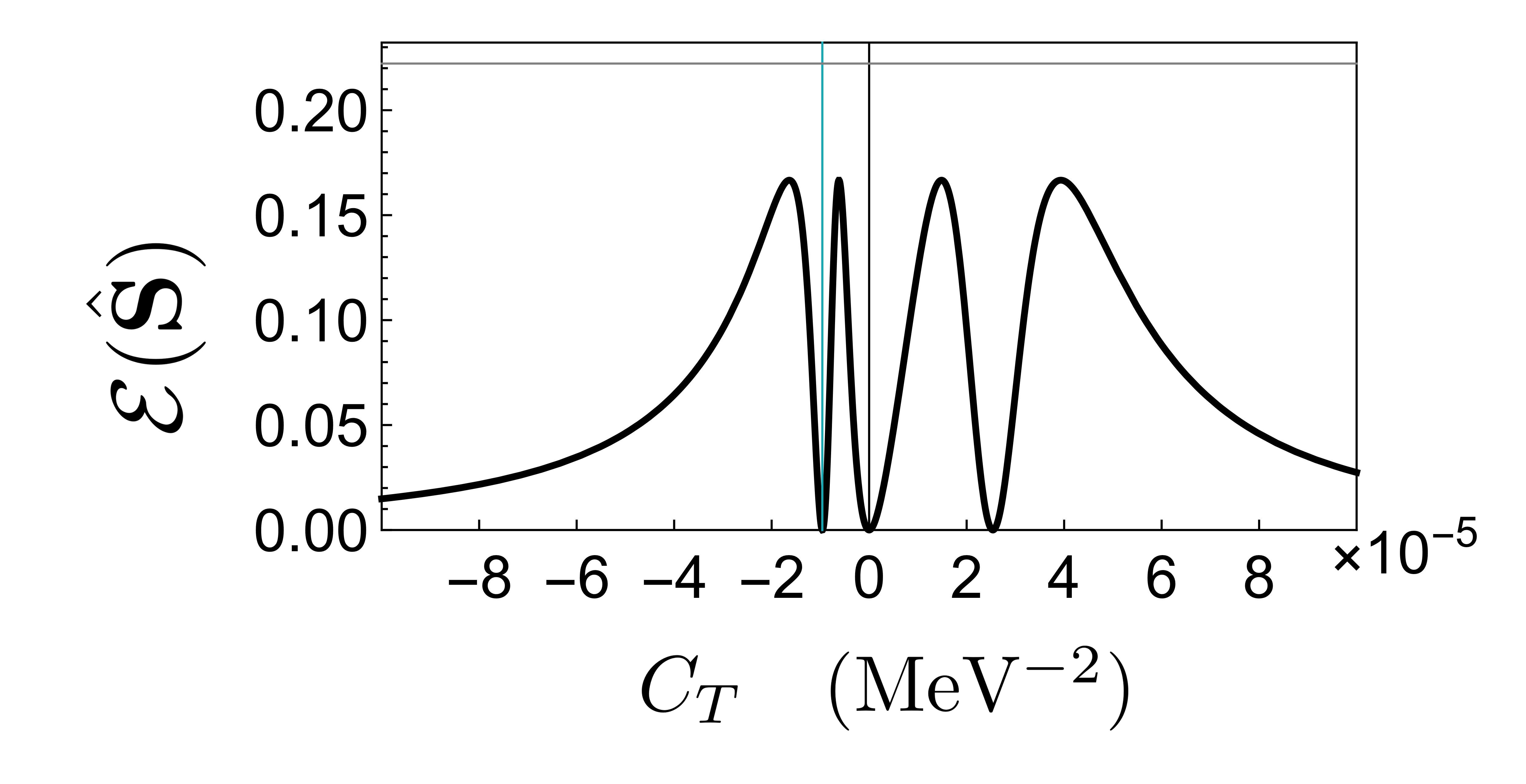}
  \caption{
  Entanglement power of the $S$-matrix, ${\mathcal E}({\hat {\bf S}})$, as a function of
  $C_T$ evaluated with $C_S = -1.2 \times 10^{-4}$ MeV$^{-2}$, $\mu$ = 140 MeV, $M$ = 939 MeV, and $p$ = 19.4 MeV.
  The vertical solid blue band indicates the physical value of $C_T = -9.605(3)\times 10^{-6}$ MeV$^{-2}$.
  The horizontal line at ${\mathcal E}({\hat {\bf S}}) = \frac{2}{9}$ indicates the maximum two-body entanglement power of an unconstrained operator in SU(4).
  }
  \label{fig:EP}
\end{figure}

While the main text has shown that the $S$-matrix entanglement power
may be written as a simple function of the difference between the $\si$ and $\siii$ phase shifts, it is enlightening to express the entanglement power directly in terms of lagrangian coefficients.
In the context of the NN pionless
EFT~\cite{Kaplan:1998tg,Kaplan:1998we,vanKolck:1998bw,Chen:1999tn},
the nucleon-nucleon scattering amplitude (not including $^3S_1\text{-}^3D_1$ mixing) may be determined in the PDS scheme~\cite{Kaplan:1998tg,Kaplan:1998we} as
\begin{equation}
  i \mathcal{A} = \frac{-i C(p^2,\mu)}{1 + MC(p^2,\mu) \left( \mu + i p\right) /4\pi} \ \ \ ,
\end{equation}
furnishing a unitary $S$-matrix,
\begin{equation}
  S = 1 + i \left(\frac{p M}{2 \pi}\right) \mathcal{A}
  \ \ ,
\end{equation}
where $M$ is the nucleon mass, $\mu$ is the renormalization scale,
$p$ is the magnitude of the center-of-mass nucleon
momentum, and $C(p^2,\mu)$ is the tree-level s-wave vertex from the NN
contact interactions.  To analyze the production of entanglement in
the spin sector, this vertex
in the $\si$-$\siii$ channels
is decomposed as
\begin{eqnarray}
C(p^2,\mu)_\sigma & = & C_S\ \hat {\bf 1}
\ +\  C_T\  \hat {\bm \sigma} \cdot \hat {\bm \sigma}
\ \ \ .
\label{eq:HLO}
\end{eqnarray}
The relation between the $\mu$-dependent coefficients and the phase shifts is
\begin{equation}
  p \cot \delta_{r} =  - \left( \frac{4 \pi}{M \overbar{C}_{r}} + \mu \right) \ \ \ ,
\end{equation}
for $r$ = 0,1 where $\delta_0 , \delta_1$ are the scattering phase shifts in the $\si$ and $\siii$ channels and spectroscopic coefficients are related to those in the vertex as
\begin{eqnarray}
\overbar{C}_{0} & = & (C_S-3 C_T) \quad
\overbar{C}_{1} = (C_S+ C_T)
\ \ \ .
\label{eq:CCdef}
\end{eqnarray}
These relations lead from the lagrangian interactions to the $S$-matrix structure of Eq.~(3) of the main text and may be expressed here as
\begin{eqnarray}
  \hat {\bf S}
  & = &
  \left[1-\frac{M p}{2}  \left[\frac{3 \overbar{C}_1}{D^-_1 }+\frac{\overbar{C}_0}{D^-_0}\right]\right]
 \hat   {\bf 1}
  + \frac{8 i \pi  C_T M p}{D^-_0D^-_1}
  \hat  {\bm \sigma} \cdot   \hat  {\bm \sigma}
  \ \ \  , \ \ \ D_j^\pm =  M \overbar{C}_j (p\pm i \mu )\pm 4 i \pi
\ \ \ .
  \label{eq:Sdef}
\end{eqnarray}
The entanglement power of the $S$-matrix becomes
\begin{equation}
 \mathcal{E}( \hat {\bf S}) =  \frac{512 \pi ^2 C_T^2 M^2 p^2 \left( 8 \pi M \mu  (C_S-C_T)+M^2 (p^2+\mu^2) \overbar{C}_1 \overbar{C}_0+16 \pi ^2\right)^2}{3 (D_0^-)^2 (D_1^-)^2 (D_0^+ )^2 (D_1^+)^2} \ \ \ .
 \label{eq:SmatEP}
\end{equation}
The curve of Fig.~\ref{fig:EP} shows this entanglement power as a function of the $\hat {\bm \sigma} \cdot \hat {\bm \sigma}$ interaction coefficient, $C_T$, focusing on the leading interaction in the effective lagrangian with the following numerical values:  $C_S =  -1.2 \times 10^{-4}$ MeV$^{-2}$, $\mu = 140$ MeV, $M = 939$ MeV and $p = 19.4$ MeV.  The following scattering parameters have been used to calculate physical values of $C_S$ and $C_T$: $a_{0} = -23.714(13)$ fm, $a_{1} = 5.425(1)$ fm, $r_{0} = 2.73(3)$ fm, and $r_{1} = 1.749(8)$ fm, where $a$ and $r$ are the scattering length and effective range, respectively.

\newpage
\setlength\LTleft{-30pt}%
\setlength\LTright{-30pt}
\begin{center}
\begin{longtable}{c|cccc}
\hline
\hline
$p$ & $ \mathcal{E}(\hat{\mathbf{S}} )_\text{pwa93} $ \ \cite{PhysRevC.48.792} & $ \mathcal{E}(\hat{\mathbf{S}})_\text{esc96} $ \ \cite{PhysRevC.54.2851,PhysRevC.54.2869} & $ \mathcal{E}(\hat{\mathbf{S}})_\text{nijm1} $ \cite{PhysRevC.49.2950} & $ \mathcal{E}(\hat{\mathbf{S}})_\text{reid93} $\ \cite{PhysRevC.49.2950} \\
\hline
\hline
0.0000 & $ 9.9984\times 10^{-33}$ & $ 9.9984\times 10^{-33}$ & $ 9.9984\times 10^{-33}$ & $ 9.9984\times 10^{-33}$ \\
0.6852 & $ 0.0067$ & $ 0.0067$ & $ 0.0003$ & $ 0.0003$ \\
1.1868 & $ 0.0255$ & $ 0.0255$ & $ 0.0031$ & $ 0.0031$ \\
1.5322 & $ 0.0312$ & $ 0.0312$ & $ 0.0083$ & $ 0.0083$ \\
1.9380 & $ 0.0446$ & $ 0.0446$ & $ 0.0204$ & $ 0.0204$ \\
2.2726 & $ 0.0635$ & $ 0.0635$ & $ 0.0363$ & $ 0.0363$ \\
2.6538 & $ 0.0837$ & $ 0.0837$ & $ 0.0604$ & $ 0.0604$ \\
2.9867 & $ 0.0977$ & $ 0.0977$ & $ 0.0845$ & $ 0.0845$ \\
3.3568 & $ 0.1101$ & $ 0.1101$ & $ 0.1100$ & $ 0.1101$ \\
3.6899 & $ 0.1209$ & $ 0.1209$ & $ 0.1284$ & $ 0.1285$ \\
3.9954 & $ 0.1311$ & $ 0.1311$ & $ 0.1407$ & $ 0.1407$ \\
4.3336 & $ 0.1419$ & $ 0.1419$ & $ 0.1497$ & $ 0.1498$ \\
4.6473 & $ 0.1506$ & $ 0.1506$ & $ 0.1551$ & $ 0.1551$ \\
4.9883 & $ 0.1578$ & $ 0.1578$ & $ 0.1591$ & $ 0.1591$ \\
5.3075 & $ 0.1624$ & $ 0.1624$ & $ 0.1622$ & $ 0.1622$ \\
5.6086 & $ 0.1651$ & $ 0.1651$ & $ 0.1648$ & $ 0.1648$ \\
5.9340 & $ 0.1665$ & $ 0.1665$ & $ 0.1665$ & $ 0.1665$ \\
6.2425 & $ 0.1665$ & $ 0.1665$ & $ 0.1665$ & $ 0.1665$ \\
6.5722 & $ 0.1653$ & $ 0.1653$ & $ 0.1653$ & $ 0.1653$ \\
6.8862 & $ 0.1629$ & $ 0.1629$ & $ 0.1629$ & $ 0.1629$ \\
7.1864 & $ 0.1597$ & $ 0.1597$ & $ 0.1596$ & $ 0.1597$ \\
7.5060 & $ 0.1554$ & $ 0.1554$ & $ 0.1553$ & $ 0.1554$ \\
7.8125 & $ 0.1506$ & $ 0.1506$ & $ 0.1504$ & $ 0.1505$ \\
8.3920 & $ 0.1402$ & $ 0.1402$ & $ 0.1398$ & $ 0.1399$ \\
8.9339 & $ 0.1292$ & $ 0.1292$ & $ 0.1288$ & $ 0.1290$ \\
9.4448 & $ 0.1182$ & $ 0.1182$ & $ 0.1180$ & $ 0.1182$ \\
9.9295 & $ 0.1075$ & $ 0.1075$ & $ 0.1075$ & $ 0.1078$ \\
10.3916 & $ 0.0973$ & $ 0.0974$ & $ 0.0974$ & $ 0.0977$ \\
10.3916 & $ 0.0973$ & $ 0.0974$ & $ 0.0974$ & $ 0.0977$ \\
11.2590 & $ 0.0789$ & $ 0.0790$ & $ 0.0790$ & $ 0.0793$ \\
12.0642 & $ 0.0632$ & $ 0.0633$ & $ 0.0631$ & $ 0.0635$ \\
13.0008 & $ 0.0469$ & $ 0.0470$ & $ 0.0468$ & $ 0.0473$ \\
13.8743 & $ 0.0340$ & $ 0.0341$ & $ 0.0339$ & $ 0.0344$ \\
14.6959 & $ 0.0239$ & $ 0.0240$ & $ 0.0238$ & $ 0.0243$ \\
15.6250 & $ 0.0149$ & $ 0.0149$ & $ 0.0148$ & $ 0.0152$ \\
16.5018 & $ 0.0085$ & $ 0.0085$ & $ 0.0084$ & $ 0.0087$ \\
17.3344 & $ 0.0042$ & $ 0.0042$ & $ 0.0041$ & $ 0.0044$ \\
18.2577 & $ 0.0012$ & $ 0.0013$ & $ 0.0012$ & $ 0.0014$ \\
19.1366 & $ 0.0001$ & $ 0.0001$ & $ 0.0001$ & $ 0.0001$ \\
19.9769 & $ 0.0002$ & $ 0.0002$ & $ 0.0003$ & $ 0.0002$ \\
20.7832 & $ 0.0014$ & $ 0.0014$ & $ 0.0015$ & $ 0.0013$ \\
21.6679 & $ 0.0038$ & $ 0.0037$ & $ 0.0038$ & $ 0.0035$ \\
22.5180 & $ 0.0068$ & $ 0.0067$ & $ 0.0069$ & $ 0.0065$ \\
23.3371 & $ 0.0105$ & $ 0.0103$ & $ 0.0106$ & $ 0.0100$ \\
24.2255 & $ 0.0150$ & $ 0.0148$ & $ 0.0152$ & $ 0.0144$ \\
25.0825 & $ 0.0199$ & $ 0.0197$ & $ 0.0201$ & $ 0.0192$ \\
25.9111 & $ 0.0250$ & $ 0.0248$ & $ 0.0252$ & $ 0.0242$ \\
26.7140 & $ 0.0302$ & $ 0.0299$ & $ 0.0304$ & $ 0.0293$ \\
27.5788 & $ 0.0361$ & $ 0.0357$ & $ 0.0363$ & $ 0.0350$ \\
28.4172 & $ 0.0419$ & $ 0.0415$ & $ 0.0421$ & $ 0.0408$ \\
29.2317 & $ 0.0476$ & $ 0.0472$ & $ 0.0479$ & $ 0.0464$ \\
33.5678 & $ 0.0779$ & $ 0.0773$ & $ 0.0782$ & $ 0.0763$ \\
36.8992 & $ 0.0991$ & $ 0.0983$ & $ 0.0995$ & $ 0.0974$ \\
39.9537 & $ 0.1160$ & $ 0.1152$ & $ 0.1164$ & $ 0.1143$ \\
42.7908 & $ 0.1292$ & $ 0.1284$ & $ 0.1296$ & $ 0.1276$ \\
45.4511 & $ 0.1395$ & $ 0.1388$ & $ 0.1399$ & $ 0.1380$ \\
47.9640 & $ 0.1475$ & $ 0.1467$ & $ 0.1478$ & $ 0.1461$ \\
50.3518 & $ 0.1535$ & $ 0.1528$ & $ 0.1538$ & $ 0.1523$ \\
52.6313 & $ 0.1580$ & $ 0.1574$ & $ 0.1583$ & $ 0.1570$ \\
54.8161 & $ 0.1614$ & $ 0.1608$ & $ 0.1616$ & $ 0.1605$ \\
56.9170 & $ 0.1637$ & $ 0.1633$ & $ 0.1639$ & $ 0.1630$ \\
58.9432 & $ 0.1653$ & $ 0.1650$ & $ 0.1654$ & $ 0.1648$ \\
60.9020 & $ 0.1662$ & $ 0.1660$ & $ 0.1663$ & $ 0.1659$ \\
62.7997 & $ 0.1666$ & $ 0.1665$ & $ 0.1666$ & $ 0.1665$ \\
64.6417 & $ 0.1666$ & $ 0.1667$ & $ 0.1666$ & $ 0.1667$ \\
66.4327 & $ 0.1663$ & $ 0.1665$ & $ 0.1662$ & $ 0.1665$ \\
68.1766 & $ 0.1657$ & $ 0.1660$ & $ 0.1656$ & $ 0.1661$ \\
69.8770 & $ 0.1649$ & $ 0.1653$ & $ 0.1647$ & $ 0.1654$ \\
71.5371 & $ 0.1639$ & $ 0.1644$ & $ 0.1637$ & $ 0.1646$ \\
73.1594 & $ 0.1628$ & $ 0.1634$ & $ 0.1625$ & $ 0.1636$ \\
74.7466 & $ 0.1615$ & $ 0.1623$ & $ 0.1612$ & $ 0.1625$ \\
76.3007 & $ 0.1602$ & $ 0.1611$ & $ 0.1599$ & $ 0.1613$ \\
77.8238 & $ 0.1588$ & $ 0.1598$ & $ 0.1584$ & $ 0.1600$ \\
89.0761 & $ 0.1465$ & $ 0.1485$ & $ 0.1459$ & $ 0.1484$ \\
100.2362 & $ 0.1331$ & $ 0.1360$ & $ 0.1323$ & $ 0.1355$ \\
110.2726 & $ 0.1214$ & $ 0.1251$ & $ 0.1205$ & $ 0.1239$ \\
120.4473 & $ 0.1104$ & $ 0.1149$ & $ 0.1094$ & $ 0.1129$ \\
130.7280 & $ 0.1002$ & $ 0.1055$ & $ 0.0993$ & $ 0.1027$ \\
141.0915 & $ 0.0911$ & $ 0.0972$ & $ 0.0902$ & $ 0.0935$ \\
151.5208 & $ 0.0830$ & $ 0.0899$ & $ 0.0821$ & $ 0.0852$ \\
162.0032 & $ 0.0757$ & $ 0.0834$ & $ 0.0750$ & $ 0.0778$ \\
172.5291 & $ 0.0692$ & $ 0.0778$ & $ 0.0687$ & $ 0.0712$ \\
183.0910 & $ 0.0635$ & $ 0.0729$ & $ 0.0632$ & $ 0.0654$ \\
193.6829 & $ 0.0584$ & $ 0.0686$ & $ 0.0583$ & $ 0.0603$ \\
203.7248 & $ 0.0541$ & $ 0.0651$ & $ 0.0543$ & $ 0.0561$ \\
213.8441 & $ 0.0502$ & $ 0.0620$ & $ 0.0506$ & $ 0.0524$ \\
224.0302 & $ 0.0466$ & $ 0.0593$ & $ 0.0474$ & $ 0.0491$ \\
234.2745 & $ 0.0435$ & $ 0.0569$ & $ 0.0445$ & $ 0.0462$ \\
244.5696 & $ 0.0406$ & $ 0.0549$ & $ 0.0419$ & $ 0.0436$ \\
254.9094 & $ 0.0380$ & $ 0.0531$ & $ 0.0395$ & $ 0.0414$ \\
265.2886 & $ 0.0357$ & $ 0.0516$ & $ 0.0374$ & $ 0.0395$ \\
275.7028 & $ 0.0336$ & $ 0.0503$ & $ 0.0356$ & $ 0.0378$ \\
285.7377 & $ 0.0318$ & $ 0.0493$ & $ 0.0340$ & $ 0.0365$ \\
295.8290 & $ 0.0302$ & $ 0.0484$ & $ 0.0325$ & $ 0.0353$ \\
305.9711 & $ 0.0288$ & $ 0.0476$ & $ 0.0312$ & $ 0.0343$ \\
316.1591 & $ 0.0276$ & $ 0.0470$ & $ 0.0300$ & $ 0.0334$ \\
326.3886 & $ 0.0265$ & $ 0.0466$ & $ 0.0289$ & $ 0.0327$ \\
336.6561 & $ 0.0256$ & $ 0.0462$ & $ 0.0280$ & $ 0.0321$ \\
346.9579 & $ 0.0248$ & $ 0.0460$ & $ 0.0271$ & $ 0.0316$ \\
356.9626 & $ 0.0243$ & $ 0.0458$ & $ 0.0264$ & $ 0.0312$ \\
367.0144 & $ 0.0239$ & $ 0.0458$ & $ 0.0257$ & $ 0.0310$ \\
377.1095 & $ 0.0236$ & $ 0.0458$ & $ 0.0250$ & $ 0.0307$ \\
387.2445 & $ 0.0235$ & $ 0.0459$ & $ 0.0245$ & $ 0.0306$ \\
\hline
\hline
\multicolumn{5}{l}{} \\[-0pt]
\caption{
    Entanglement power, $\mathcal{E}(\hat{\mathbf{S}})$, of the
    $S$-matrix as a function of $p$, the center-of-mass nucleon momentum.  This data is graphically shown in Fig.~1 of the main text.  The $\si$ and $\siii$ phase shifts used to calculate $\mathcal{E}(\hat{\mathbf{S}})$ were determined by four different models accessed through the NN-Online database~\cite{NNOnline}.
    }
    \label{tab:fig1mt}
\end{longtable}
\end{center}

\end{document}